\documentclass[journal]{IEEEtran}

\usepackage{amssymb}

\usepackage{cite}

\usepackage{graphicx}

\usepackage[caption=false,font=footnotesize]{subfig}

\usepackage{amsmath}
\usepackage{algorithm}
\usepackage{algorithmic}
\usepackage{multirow}

\usepackage{float}

\begin{document}
%
% paper title
% Titles are generally capitalized except for words such as a, an, and, as,
% at, but, by, for, in, nor, of, on, or, the, to and up, which are usually
% not capitalized unless they are the first or last word of the title.
% Linebreaks \\ can be used within to get better formatting as desired.
% Do not put math or special symbols in the title.
\title{Adaptive Traffic Signal Control: Deep Reinforcement Learning Algorithm with Experience Replay and Target Network}
%\title{Maximizing User Rewards in Incentive WiFi offloading with Energy Consumption Constraint}

 %author names and affiliations
 %use a multiple column layout for up to three different
 %affiliations
%\author{\IEEEauthorblockN{Juntao Gao}
%\IEEEauthorblockA{Graduate School of Information Science\\
%Nara Institute of Science and Technology, Nara, Japan\\
%Email: \{jtgao\}@is.naist.jp}
%\IEEEauthorblockN{Minoru Ito}
%\IEEEauthorblockA{Graduate School of Information Science\\
%Nara Institute of Science and Technology, Nara, Japan\\
%Email: \{ito\}@is.naist.jp}
%
%}
\author{Juntao Gao,
				Yulong Shen,
        Jia Liu,
				Minoru Ito
        and Norio Shiratori
\thanks{J. Gao and M. Ito are with the Graduate School of Information Science, Nara Institute of Science and Technology, 8916-5 Takayama-cho, Ikoma, Nara 630-0192, JAPAN. E-mail: \{jtgao,ito\}@is.naist.jp.}% <-this % stops a space
\thanks{Y. Shen is with School of Computer Science and Technology, Xidian University, Xian, Shaanxi 710071, PR China. E-mail: ylshen@mail.xidian.edu.cn.}
\thanks{J. Liu is with National Institute of Informatics, JAPAN.}
\thanks{N. Shiratori is with Tohoku University, Sendai, JAPAN.}

}

% make the title area
\maketitle

\begin{abstract}
Adaptive traffic signal control, which adjusts traffic signal timing according to real-time traffic, has been shown to be an effective method to reduce traffic congestion.
Available works on adaptive traffic signal control make responsive traffic signal control decisions based on human-crafted features (e.g. vehicle queue length).
%pre-processed raw real-time traffic data (e.g. position and speed of vehicles) to extract handcrafted features (e.g. vehicle queue length and average vehicle delay) to make responsive traffic signal control decisions.
%To make wise responsive traffic signal control decisions, a control agent usually uses handcrafted features, such as vehicle queue length and average vehicle delay, to represent real-time traffic situation.
%However, human-crafted features ignore some useful traffic information (e.g. vehicle queue length does not consider vehicles not in queue but to come soon), leading to suboptimal traffic signal controls.
However, human-crafted features are abstractions of raw traffic data (e.g., position and speed of vehicles), which ignore some useful traffic information and lead to suboptimal traffic signal controls.
In this paper, we propose a deep reinforcement learning algorithm that automatically extracts all useful features (machine-crafted features) from raw real-time traffic data and learns the optimal policy for adaptive traffic signal control. 
To improve algorithm stability, we adopt experience replay and target network mechanisms.
Simulation results show that our algorithm reduces vehicle delay by up to $47\%$ and $86\%$ when compared to another two popular traffic signal control algorithms, longest queue first algorithm and fixed time control algorithm, respectively.
\end{abstract}

\IEEEpeerreviewmaketitle

%%%%%%%%%%%%%%%%%%%%%%%%%%%%%%%%%%%%%%%%%%%%%%%%%%%%%%%%%%%%%%%%%%%%%%%%%%%%%%%%%%%%%%%%%%
%%%%%%%%%%%%%%%%%%%%%%%%%%%%%%%%%%%%%%%%%%%%%%%%%%%%%%%%%%%%%%%%%%%%%%%%%%%%%%%%%%%%%%%%%%
\section{Introduction}
Traffic congestion has led to some serious social problems: long travelling time, fuel consumption, air pollution, etc \cite{Zhao_SMC12,Alsabaan_CST13}.  Factors responsible for traffic congestion include: proliferation of vehicles, inadequate traffic infrastructure and inefficient traffic signal control. However, we cannot stop people from buying vehicles and building new traffic infrastructure is of high cost. The relatively easy solution is to improve efficiency of traffic signal control. Fixed-time traffic signal control is common in use, where traffic signal timing at an intersection is predetermined and optimized offline based on history traffic data (not real-time traffic demands). However, traffic demands may change from time to time, making predetermined settings of traffic signal timing out of date. Therefore, fixed-time traffic signal control cannot adapt to dynamic and bursty traffic demands, resulting in traffic congestion.

%we can improve efficiency of traffic signal control relatively easily due to advances in vehicular sensing and communication technologies \cite{Zhao_SMC12,Alsabaan_CST13}. Fixed-time traffic signal control is common in use, where traffic signal timing is optimized offline based on history traffic data at an intersection. However, traffic distribution may change from time to time, making predetermined settings of traffic signal timing out of date. Therefore, fixed-time traffic signal control cannot react to dynamic and bursty traffic demands, resulting in traffic congestion.

In contrast, adaptive traffic signal control, which adjusts traffic signal timing according to real-time traffic demand, has been shown to be an effective method to reduce traffic congestion \cite{Zaidi_ITS16,Gregoire_CNS15,LA_ITS11,Yin_CDC15,Arel_IET10,Mannion_chapter16}. For example, Zaidi \textit{et al.} \cite{Zaidi_ITS16} and  Gregoire \textit{et al.} \cite{Gregoire_CNS15} proposed adaptive traffic signal control algorithms based on back-pressure method, which is similar to pushing water (here vehicles) to flow through a network of pipes (roads) by pressure gradients (the number of queued vehicles) \cite{Neely_PhD03}. Authors in \cite{LA_ITS11,Yin_CDC15,Arel_IET10,Mannion_chapter16} proposed to use reinforcement learning method to adaptively control traffic signals, where they modelled the control problem as a Markov decision process \cite{Sutton_RL98}. However, all these works make responsive traffic signal control decisions based on human-crafted features, such as vehicle queue length and average vehicle delay. Human-crafted features are abstractions of raw traffic data (e.g., position and speed of vehicles), which ignore some useful traffic information and lead to suboptimal traffic signal controls. For example, vehicle queue length does not consider vehicles that are not in queue but will come soon, which is also useful information for controlling traffic signals; average vehicle delay only reflects history traffic data not real-time traffic demand. 

In this paper, instead of using human-crafted features, we propose a deep reinforcement learning algorithm that automatically extracts all features (machine-crafted features) useful for adaptive traffic signal control from raw real-time traffic data and learns the optimal traffic signal control policy. Specifically, we model the control problem as a reinforcement learning problem \cite{Sutton_RL98}. Then, we use deep convolutional neural network to extract useful features from raw real-time traffic data (i.e., vehicle position, speed and traffic signal state) and output the optimal traffic signal control decision.
A well-known problem with deep reinforcement learning is that the algorithm may be unstable or even diverge in decision making \cite{Mnih_Nature15}. To improve algorithm stability, we adopt two methods proposed in \cite{Mnih_Nature15}: experience replay and target network (see details in Section \ref{section:drl}). 

The rest of this paper is organized as follows. In Section \ref{section:pf}, we introduce intersection model and define reinforcement learning components: intersection state, agent action, reward and agent goal. In Section \ref{section:drl}, we present details of our proposed deep reinforcement learning algorithm for traffic signal control. In Section \ref{section:ev}, we verify our algorithm by simulations and compare its performance to popular traffic signal control algorithms. In Section \ref{section:rw}, we review related work on adopting deep reinforcement learning for traffic signal control and their limitations and conclude the whole paper in Section \ref{section:conclusion}.

\begin{figure}[!th]
\centering
\includegraphics[width=3.1in]{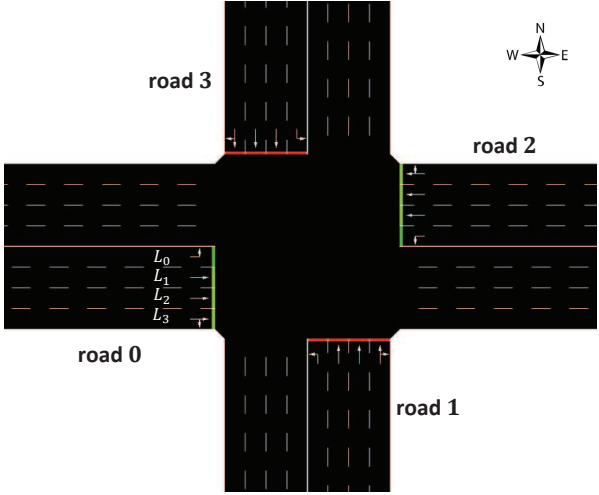}
\caption{A four-way intersection.}
\label{fig:intersection}
\end{figure}

%%%%%%%%%%%%%%%%%%%%%%%%%%%%%%%%%%%%%%%%%%%%%%%%%%%%%%%%%%%%%%%%%%%%%%%%%%%%%%%%%%%%%%%%%%
%%%%%%%%%%%%%%%%%%%%%%%%%%%%%%%%%%%%%%%%%%%%%%%%%%%%%%%%%%%%%%%%%%%%%%%%%%%%%%%%%%%%%%%%%%
\section{System Model and Problem Formulation} \label{section:pf}
In this section, we first introduce intersection model and then formulate traffic signal control problem as a reinforcement learning problem.

Consider a four-way intersection in Fig.\ref{fig:intersection}, where each road consists of four lanes. For each road, the innermost lane (referred to as $L_0$) is for vehicles turning left, the middle two lanes ($L_1$ and $L_2$) are for vehicles going straight and the outermost lane ($L_3$) is for vehicles going straight or turning right. Vehicles at this intersection run under control of traffic signals: green lights mean vehicles can go through the intersection, however vehicles at left-turn waiting area should let vehicles going straight pass first; yellow lights mean lights are about to turn red and vehicles should stop if it is safe to do so; red lights mean vehicles must stop. For example, green lights for west-east traffic are turned on in Fig.\ref{fig:intersection}.  

\begin{figure}[!t]
\centering
\includegraphics[width=3.3in]{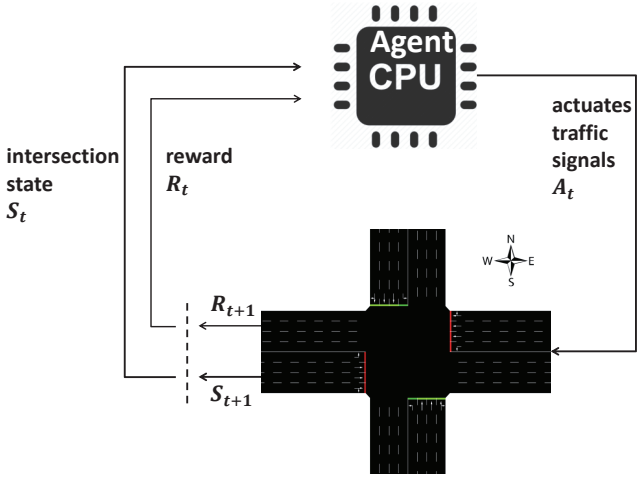}
\caption{Reinforcement learning agent for traffic signal control.}
\label{fig:agent}
\end{figure}

We formulate traffic signal control problem as a reinforcement learning problem shown in Fig.\ref{fig:agent} \cite{Sutton_RL98}, where an agent interacts with the intersection at discrete time steps, $t =0,1,2, \cdots$, and the goal of the agent is to reduce vehicle staying time at this intersection in the long run, thus alleviating traffic congestion. Specifically, such an agent first observes intersection state $S_t$ (defined later) at the beginning of time step $t$, then selects and actuates traffic signals $A_t$. After vehicles move under actuated traffic signals, intersection state changes to a new state $S_{t+1}$. The agent also gets reward $R_{t}$ (defined later) at the end of time step $t$ as a consequence of its decision on selecting traffic signals. Such reward serves as a signal guiding the agent to achieve its goal. In time sequence, the agent interacts with the intersection as $\cdots, S_t, A_t, R_{t}, S_{t+1}, A_{t+1} \cdots$. Next, we define intersection state $S_t$, agent action $A_t$ and reward $R_{t}$, respectively.

\begin{figure}[!t]
\centering
\includegraphics[width=3.3in]{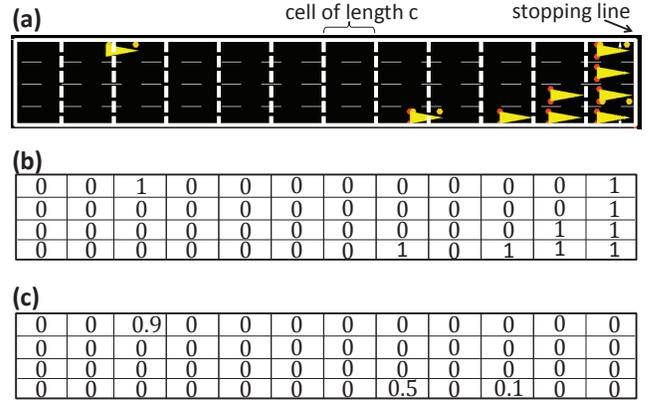}
\caption{$(a)$ snapshot of traffic at road 0. $(b)$ matrix of vehicle position. $(c)$ matrix of normalized vehicle speed. }
\label{fig:state}
\end{figure}

\textbf{Intersection State:} Intersection information needed by the agent to control traffic signals includes vehicle position, vehicle speed at each road and traffic signal state. To easily represent information of vehicle position and vehicle speed (following methods in \cite{Genders16}), we divide lane segment of length $l$, starting from stop line, into discrete cells of length $c$ for each road $i=0,1,2,3$ as illustrated in Fig. \ref{fig:state}. We then collect vehicle position and speed information of road $i$ into two matrices: matrix of vehicle position $\mathbf{P}_i$ and matrix of vehicle speed $\mathbf{V}_i$. If a vehicle is present at one cell, the corresponding entry of matrix $\mathbf{P}_i$ is set to $1$. The vehicle speed, normalized by road speed limit, is recorded at the corresponding entry of matrix $\mathbf{V}_i$. The matrix $\mathbf{P}$ of vehicle position for all roads of the intersection is then given by
\begin{align} \label{eq:position}
\mathbf{P}=& \left[\begin{array}{c}
										\mathbf{P}_0 \\
										\mathbf{P}_2 \\
										\mathbf{P}_1 \\
										\mathbf{P}_3 
										\end{array}
						\right]
\end{align}
Similarly, the matrix $\mathbf{V}$ of vehicle speed for all roads of the intersection is given by
\begin{align} \label{eq:speed}
\mathbf{V}=& \left[\begin{array}{c}
										\mathbf{V}_0 \\
										\mathbf{V}_2 \\
										\mathbf{V}_1 \\
										\mathbf{V}_3 
										\end{array}
						\right]
\end{align}

To represent the state of selected traffic signals, we use a vector $\mathbf{L}$ of size $2$ since the agent can only choose between two actions: turning on green lights for west-east traffic (i.e., red lights for north-south traffic) or turning on green lights for north-south traffic (i.e., red lights for west-east traffic). When green lights are turned on for west-east traffic, $\mathbf{L}=[1, 0]$; when green lights are turned on for north-south traffic, $\mathbf{L}=[0, 1]$.

In summary, at the beginning of time step $t$, the agent observes intersection state $S_t=(\mathbf{P}, \mathbf{V}, \mathbf{L}) \in \mathcal{S}$ for traffic signal control, where $\mathcal{S}$ denotes the whole state space.

\begin{figure}[!t]
\centering
\includegraphics[width=3.3in]{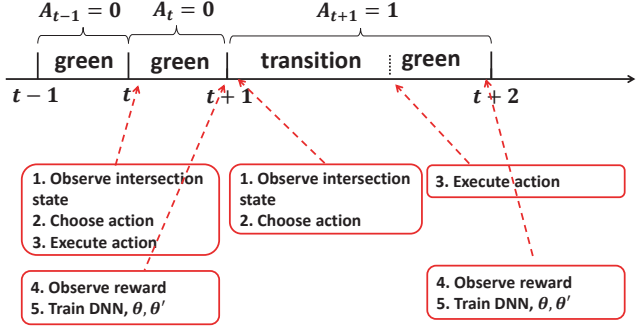}
\caption{Timeline for agent events: observation, choosing action, executing action, observing rewards and training DNN network.}
\label{fig:timeline}
\end{figure}

\begin{figure}[!t]
\centering
\includegraphics[width=3.5in]{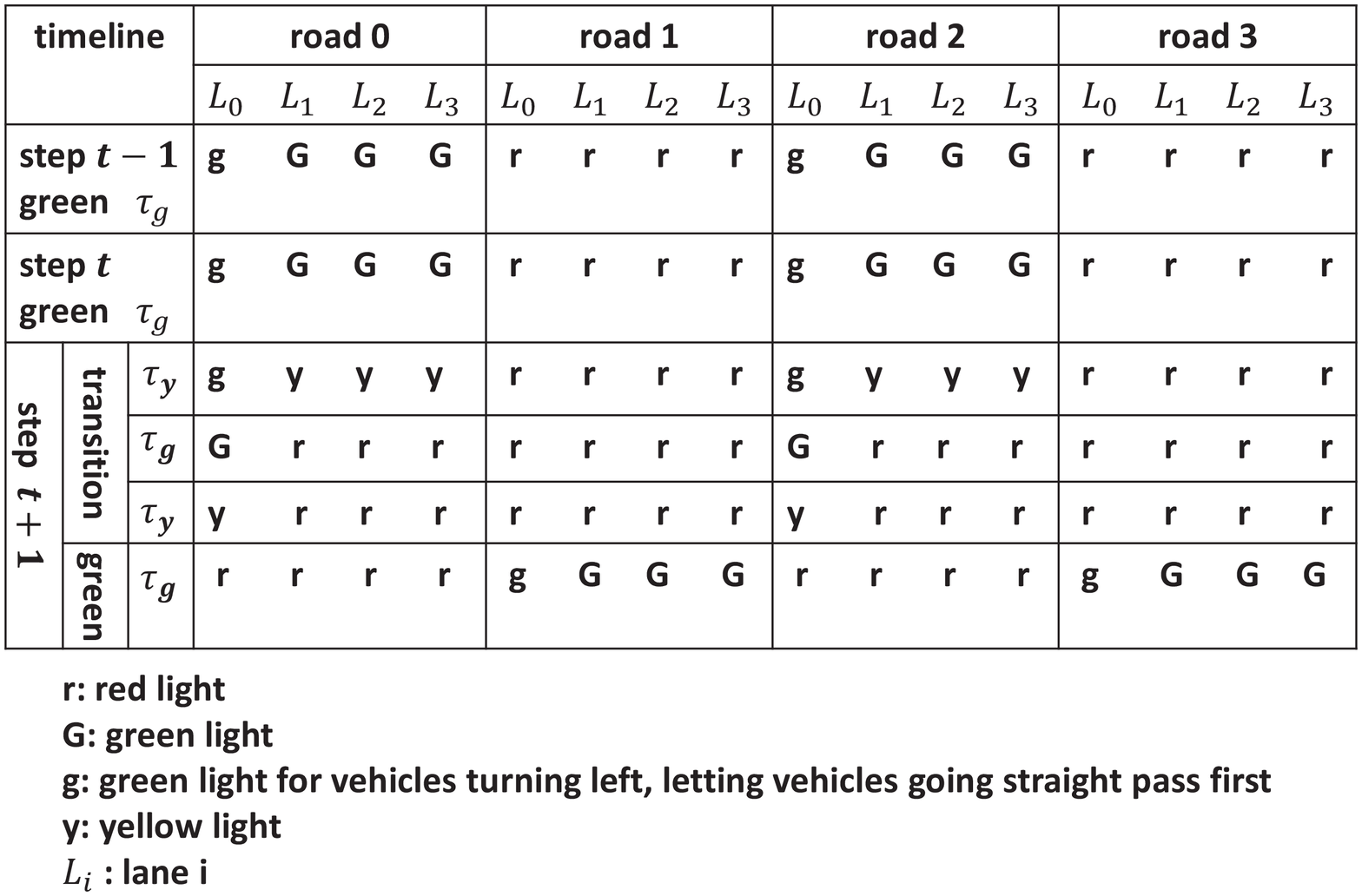}
\caption{Example of traffic signal timing for actions in Fig. \ref{fig:timeline}.}
\label{fig:signaltiming}
\end{figure}

\textbf{Agent Action:} As shown in Fig.\ref{fig:timeline}, after observing intersection state $S_t$ at the beginning of each time step $t$, the agent chooses one action $A_t \in \mathcal{A}, \mathcal{A}=\{0,1\}$: turning on green lights for west-east traffic ($A_t=0$) or for north-south traffic ($A_t=1$), and then executes the chosen action. Green lights for each action last for fixed time interval of length $\tau_g$. When green light interval ends, the current time step $t$ ends and new time step $t+1$ begins. The agent then observes new intersection state $S_{t+1}$ and chooses the next action $A_{t+1}$ (the same action may be chosen consecutively across time steps, e.g., steps $t-1$ and $t$ in Fig.\ref{fig:timeline}). If the chosen action $A_{t+1}$ at time step $t+1$ is the same with previous action $A_{t}$, simply keep current traffic signal settings unchanged. If the chosen action $A_{t+1}$ is different from previous action $A_{t}$, before the selected action $A_{t+1}$ is executed, the following transition traffic signals are actuated to clear vehicles going straight and vehicles at left-turn waiting area.
First, turn on yellow lights for vehicles going straight. All yellow lights last for fixed time interval of length $\tau_y$. Then, turn on green lights of duration $\tau_g$ for left-turn vehicles. Finally, turn on yellow lights for left-turn vehicles.
An example in Fig. \ref{fig:signaltiming} shows the traffic signal timing corresponding to the chosen actions in Fig. \ref{fig:timeline}.

Define action policy $\pi$ as rules the agent follows to choose actions after observing intersection state. For example, $\pi$ can be a random policy such that the agent chooses actions with probability $P\{A_t=a|S_t=s\}, a \in \mathcal{A}, s \in \mathcal{S}$.

\begin{figure}[!t]
\centering
\includegraphics[width=3.5in]{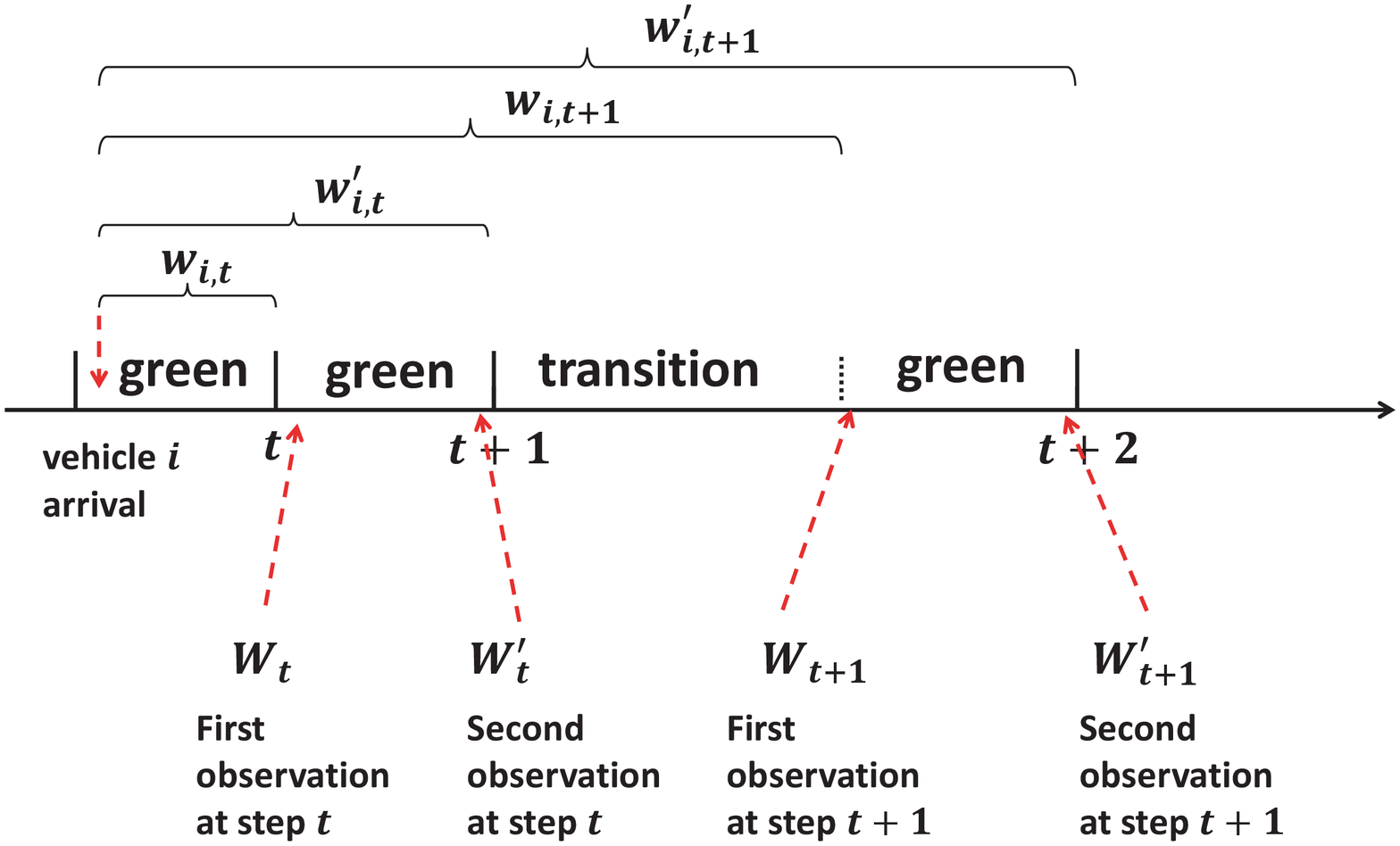}
\caption{Example of vehicle staying time.}
\label{fig:stayingtime}
\end{figure}

\textbf{Reward:} To reduce traffic congestion, it is reasonable to reward the agent at each time step for choosing some action if the time of vehicles staying at the intersection decreases. 
Specifically, the agent observes vehicles staying time twice every time step to determine its change as shown in Fig. \ref{fig:stayingtime}. The first observation is at the beginning of green light interval at each time step and the second observation is at the end of green light interval at each time step.

Let $w_{i,t}$ be the staying time (in seconds) of vehicle $i$ from the time the vehicle enters one road of the intersection to the beginning of green light interval at time step $t$ (vehicle $i$ should still be at the intersection, otherwise $w_{i,t}=0$), and $w'_{i,t}$ be the staying time of vehicle $i$ from the time the vehicle enters one road of the intersection to the end of green light interval at time step $t$.
%Define the staying time of a vehicle at an intersection as the time interval (in seconds) from the time the vehicle enters one road of the intersection to the time the agent observes its staying time. 
%Define the delay of a vehicle at an intersection as the time interval (in seconds) from the time the vehicle enters one road of the intersection to the time it leaves the intersection.
%for which the vehicle either runs on a road or waits in a queue of the intersection. 
%It is reasonable to reward the agent for choosing some action if vehicle waiting time decreases at each time step. Define the waiting time of a vehicle as the time interval (in seconds) from the time the vehicle's speed is lower than $0.1$ m/s to the time its speed changes to be higher than $0.1$ m/s. 
Similarly, let $W_t=\sum_{i}w_{i,t}$ be the sum of staying time of all vehicles at the beginning of green light interval at time step $t$, and $W'_t=\sum_{i}w'_{i,t}$ be the sum of staying time of all vehicles at the end of green light interval at time step $t$.
For example, at time step $t$ in Fig. \ref{fig:stayingtime}, $W_t$ is observed at the beginning of time step $t$ because green light interval starts at the beginning of time step $t$ and $W'_t$ is observed at the end of time step $t$. However, at time step $t+1$, $W_{t+1}$ is observed not at the beginning of time step $t+1$ but when transition interval ends and green light interval begins, $W'_{t+1}$ is observed at the end of time step $t+1$, i.e., when green light interval ends.
%Assume the agent has access to the sum $W_t$ of waiting time of all vehicles at the intersection at the beginning of time step $t$ (e.g., vehicles report their waiting time to the agent). 
At time step $t$, if the staying time $W'_{t}$ decreases, $W'_{t}<W_{t}$, the agent should be rewarded; if the staying time $W'_{t}$ increases, $W'_{t}>W_{t}$, the agent should be penalized. Thus, we define the reward $R_{t}$ for the agent choosing some action at time step $t$ as follows
\begin{align}
R_{t}=W_{t}-W'_{t} \label{eq:reward}
\end{align}

\textbf{Agent Goal:} Recall that the goal of the agent is to reduce vehicle staying time at the intersection in the long run. Suppose the agent observes intersection state $S_t$ at the beginning of time step $t$, then makes action decisions according to some action policy $\pi$ hereafter, and receives a sequence of rewards after time step $t$, $R_t$, $R_{t+1}, R_{t+2}, R_{t+3}, \cdots$. If the agent aims to reduce vehicle staying time at the intersection for one time step $t$, it is sufficient for the agent to choose one action that maximizes the immediate reward $R_{t}$ as defined in (\ref{eq:reward}). Since the agent aims to reduce vehicle staying time in the long run, the agent needs to find an action policy $\pi^{\ast}$ that maximizes the following cumulative future reward, namely Q-value, 
\begin{align}
&Q_{\pi}(s,a)\!=\!\mathbb{E}\big\{R_{t}\!+\! \gamma R_{t+1} \!+\! \gamma^2 R_{t+2} + \cdots | S_t\!=\!s, A_t\!=\!a, \pi \big\} \nonumber \\
&	\quad\quad\quad \,\, =\mathbb{E}\big\{\sum_{k=0}^{\infty}\gamma^kR_{t+k}| S_t=s, A_t=a, \pi \big\}
\end{align}
where the expectation is with respect to action policy $\pi$, $\gamma$ is a discount parameter, $0 \leq \gamma \leq 1$, reflecting how much weight the agent puts on future rewards: $\gamma=0$ means the agent is shortsighted, only considering immediate reward $R_{t}$ and $\gamma$ approaching $1$ means the agent is more farsighted, considering future rewards more heavily.

More formally, the agent needs to find an action policy $\pi^{\ast}$ such that
\begin{align}
&\pi^{\ast}=\arg\max_{\pi}Q_{\pi}(s,a) \\
& \quad \quad \quad\quad\quad \text{for all} \, s \in \mathcal{S}, a \in \mathcal{A} \nonumber
\end{align}
Denote the optimal Q-values under action policy $\pi^{\ast}$ by $Q^{\ast}(s,a)=Q_{\pi^{\ast}}(s,a)$.

%%%%%%%%%%%%%%%%%%%%%%%%%%%%%%%%%%%%%%%%%%%%%%%%%%%%%%%%%%%%%%%%%%%%%%%%%%%%%%%%%%%%%%%%%%
%%%%%%%%%%%%%%%%%%%%%%%%%%%%%%%%%%%%%%%%%%%%%%%%%%%%%%%%%%%%%%%%%%%%%%%%%%%%%%%%%%%%%%%%%%
\section{Deep Reinforcement Learning Algorithm for Traffic Signal Control} \label{section:drl}
In this section, we introduce deep reinforcement learning algorithm that extracts useful features from raw traffic data and finds the optimal traffic signal control policy $\pi^{\ast}$, and experience replay and target network mechanisms to improve algorithm stability.

%\textbf{Reinforcement Learning Problem:} If a problem can be formulated as an agent learning from interacting with environment to achieve a goal, it is called a reinforcement learning problem. Specifically, such an agent first observes environment state, then selects actions to influence environment 

If the agent already knows the optimal Q-values $Q^{\ast}(s,a)$ for all state-action pairs $s \in \mathcal{S}, a \in \mathcal{A}$, the optimal action policy $\pi^{\ast}$ is simply choosing the action $a$ that achieves the optimal value $Q^{\ast}(s,a)$ under intersection state $s$. Therefore, the agent needs to find optimal Q-values $Q^{\ast}(s,a)$ next. For  optimal Q-values $Q^{\ast}(s,a)$, we have the following recursive relationship, known as Bellman optimality equation \cite{Sutton_RL98}, 
\begin{align}
&Q^{\ast}(s,a)=\mathbb{E} \big\{R_{t}\!+\!\gamma\max_{a'}Q^{\ast}(S_{t+1},a') | S_t\!=\!s, A_t\!=\!a \big\}  \label{eq:recursive-q-value} \nonumber \\
& \quad \quad \quad\quad\quad \text{for all} \, s \in \mathcal{S}, a \in \mathcal{A}
\end{align}
The intuition is that the optimal cumulative future reward the agent receives is equal to the immediate reward it receives after choosing action $a$ at intersection state $s$ plus the optimal future reward thereafter. In principle, we can solve $(\ref{eq:recursive-q-value})$ to get optimal Q-values $Q^{\ast}(s,a)$ if the number of total states is finite and we know all details of the underlying system model, such as transition probabilities of intersection states and corresponding expected reward. However, it is too difficult, if not impossible, to get these information in reality. Complex traffic situations at the intersection constitute enormous intersection states, making it hard to find transition probabilities for those states. 

Instead of solving $(\ref{eq:recursive-q-value})$ directly, we resort to approximating those optimal Q-values $Q^{\ast}(s,a)$ by a parameterized deep neural network (DNN) such that the output of the neural network $Q(s,a;\theta)\approx Q^{\ast}(s,a)$, where $\theta$ are features/parameters that will be learned from raw traffic data. 

\begin{figure*}[!th]
\centering
\includegraphics[width=4.3in]{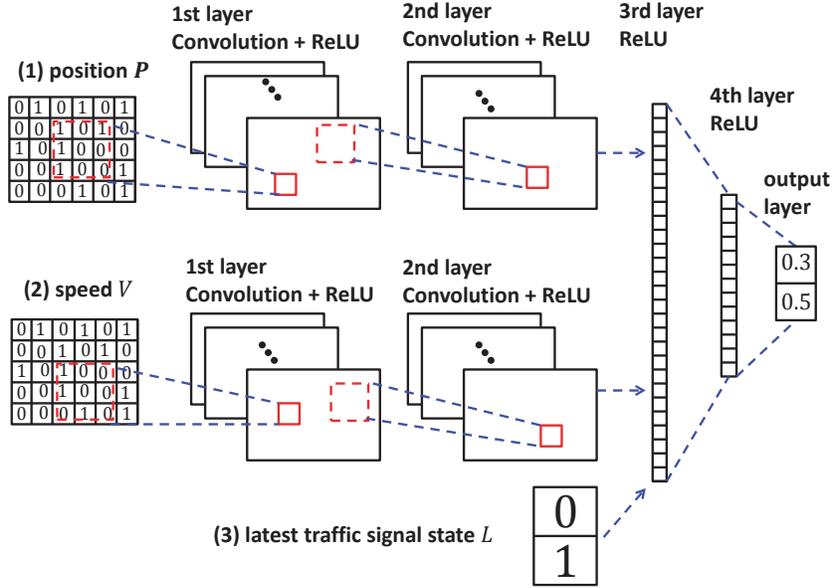}
\caption{DNN structure. Note that the small matrices and vectors in this figure are for illustration simplicity, whose dimensions should be set accordingly in DNN implementation.}
\label{fig:dnnstructure}
\end{figure*}

%\begin{figure*}[!th]
    %\centering
    %{
    %\subfloat[DNN structure.]
    %{\includegraphics[width=3.5in]{dnnstructure} \label{fig:dnnstructure}}
    %\subfloat[Agent training.]
    %{\includegraphics[width=4.5in]{training} \label{fig:training}}
    %}
    %\caption{Average vehicle delay for separate roads at the intersection under different traffic signal control algorithms.}
    %\label{fig:DNNtraining}
%\end{figure*}

\textbf{DNN Structure:} We construct such a DNN network, following the approach in \cite{Mnih_Nature15} and \cite{Genders16}, where the network input is the observed intersection state $S_t=(\mathbf{P}, \mathbf{V}, \mathbf{L})$ and the output is a vector of estimated Q-values $Q(S_t,a;\theta)$ for all actions $a \in \mathcal{A}$ under observed state $S_t$. Detailed architecture of the DNN network is given in Fig. \ref{fig:dnnstructure}: (1) position matrix $\mathbf{P}$ is fed to a stacked sub-network where the first layer convolves matrix $\mathbf{P}$ with 16 filters of $4\times 4$ with stride 2 and applies a rectifier nonlinearity activation function (ReLU), the second layer convolves the first layer output with 32 filters of $2\times 2$ with stride 1 and also applies ReLU; (2) speed matrix $\mathbf{V}$ is fed to another stacked sub-network which has the same structure with the previous sub-network, however with different parameters; (3) traffic signal state vector $\mathbf{L}$ is concatenated with the flattened outputs of the two sub-networks, forming the input of the third layer in Fig. \ref{fig:dnnstructure}. The third and fourth layers are fully connected layers of 128 and 64 units, respectively, followed by rectifier nonlinearity activation functions (ReLU). The final output layer is fully connected linear layer outputting a vector of Q-values, where each vector entry corresponds to the estimated Q-value $Q(S_t,a;\theta)$ for an action $a \in \mathcal{A}$ under state $S_t$.

\begin{algorithm}[!t]
\caption{Deep reinforcement learning algorithm with experience replay and target network for traffic signal control}
\label{algorithm:drl}
\begin{algorithmic}[1]

\STATE Initialize DNN network with random weights $\theta$;
\STATE Initialize target network with weights $\theta'=\theta$;
\STATE Initialize $\epsilon,\gamma,  \beta, N$;
\FOR {episode$=1$ to $N$}
		\STATE Initialize intersection state $S_1$;
		\STATE Initialize action $A_0$;
		\STATE Start new time step;
		\FOR {time $=1$ to $T$ seconds}
				\IF {new time step $t$ begins}
						\STATE The agent observes current intersection state $S_{t}$;
						\STATE The agent selects action $A_t=\arg\max_{a}Q(S_t,a;\theta)$ with probability $1-\epsilon$ and randomly selects an action $A_t$ with probability $\epsilon$;
						\IF{ $A_t==A_{t-1}$}
								\STATE Keep current traffic signal settings unchanged;
						\ELSE
								\STATE Actuate transition traffic signals;
						\ENDIF
						
				\ENDIF
				\STATE Vehicles run under current traffic signals;
				\STATE $time = time+1$;
				
				\IF{transition signals are actuated and transition interval ends}
						\STATE Execute selected action $A_t$;
				\ENDIF
				
				\IF{time step $t$ ends}
						\STATE The agent observes reward $R_{t}$ and current intersection state $S_{t+1}$;
						\STATE Store observed experience $(S_t, A_t, R_{t}, S_{t+1})$ into replay memory $\mathbf{M}$;
						\STATE Randomly draw $32$ samples $(S_i, A_i, R_{i}, S_{i+1})$ as minibatch from memory $\mathbf{M}$;
						\STATE Form training data: input data set $\mathbf{X}$ and targets $\mathbf{y}$;
						\STATE Update $\theta$ by applying RMSProp algorithm to training data;
						\STATE Update $\theta'$ according to ($\ref{eq:soft-update}$);
				\ENDIF
		\ENDFOR
\ENDFOR

\end{algorithmic}
\end{algorithm}

\begin{figure*}[!th]
\centering
\includegraphics[width=4.3in]{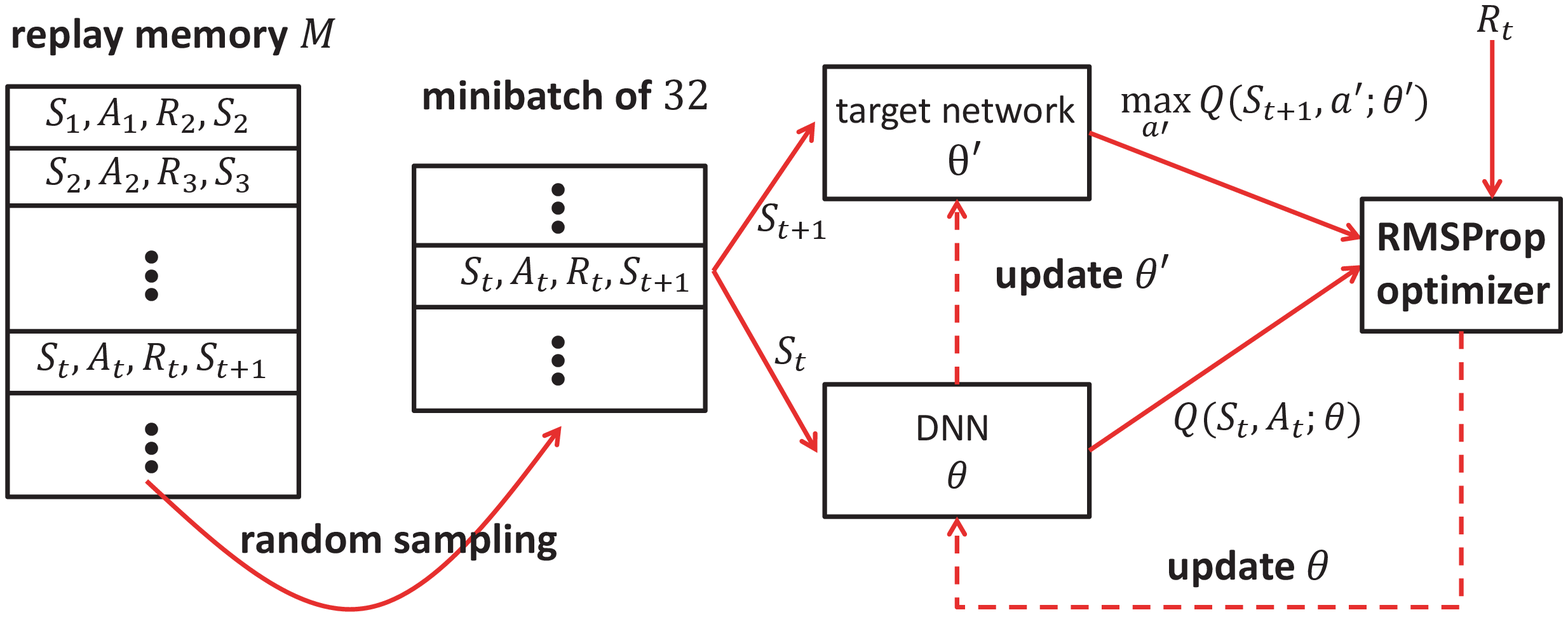}
\caption{Agent training process.}
\label{fig:training}
\end{figure*}

\textbf{DNN Training:} The whole training algorithm is summarized in Algorithm \ref{algorithm:drl} and illustrated in Fig. \ref{fig:training}. Note that time at line $8$ simulates the real world time in seconds, time step at line $9$ is one period during which agent events occur as shown in Fig. \ref{fig:timeline}. At each time step $t$, the agent records observed interaction experience $E_t=(S_t, A_t, R_{t}, S_{t+1})$ into a replay memory $\mathbf{M}=\{E_1, E_2, \cdots, E_t\}$. The replay memory is of finite capacity and when it is full, the oldest data will be discarded. To learn DNN features/parameters $\theta$ such that outputs $Q(s,a;\theta)$ best approximate $Q^{\ast}(s,a)$, the agent needs training data: input data set $\mathbf{X}=\{(S_t,A_t): t\geq 1\}$ and the corresponding targets $\mathbf{y}=\{Q^{\ast}(S_t,A_t): t\geq 1\}$. For input data set, $(S_t,A_t)$ can be retrieved from replay memory $\mathbf{M}$. However, target $Q^{\ast}(S_t,A_t)$ is not known. As in \cite{Mnih_Nature15}, we use its estimate value $R_{t}+\gamma\max_{a'}Q(S_{t+1},a';\theta')$ as the target instead, where $Q(S_{t+1},a';\theta')$ is the output of a separate target network with parameters $\theta'$ as shown in Fig. \ref{fig:training} (see $(\ref{eq:soft-update})$ for how to set $\theta'$) and the input of the target network is the corresponding $S_{t+1}$ from interaction experience $E_t=(S_t, A_t, R_{t}, S_{t+1})$. Define $Q(S_{t+1},a';\theta')=0$ if training episode terminates at time step $t+1$. The target network has the same architecture with the DNN network shown in Fig. \ref{fig:dnnstructure}. Thus, targets $\mathbf{y}=\{R_{t}+\gamma\max_{a'}Q(S_{t+1},a';\theta'): t\geq 1\}$. 

After collecting training data, the agent learns features/parameters $\theta$ by training the DNN network to minimize the following mean squared error (MSE) 
\begin{align} 
MSE(\theta)&=\frac{1}{m}\sum_{t=1}^{m}\Big\{\big(R_{t}+\gamma\max_{a'}Q(S_{t+1},a';\theta')\big) \nonumber \\
& \quad \quad\quad\quad\quad-Q(S_t,A_t;\theta)\Big\}^2
\end{align}
where $m$ is the size of input data set $\mathbf{X}$. However, if $m$ is large, the computational cost for minimizing $MSE(\theta)$ is high.
To reduce computational cost, we adopt the stochastic gradient descent algorithm RMSProp \cite{Hinton_RMSProp12} with minibatch of size $32$. Following this method, when the agent trains the DNN network, it randomly draws $32$ samples from the replay memory $\mathbf{M}$ to form $32$ input data and target pairs (referred to as experience replay), and then uses these $32$ input data and targets to update DNN parameters/features $\theta$ by RMSProp algorithm.

%minimizing 
%\begin{align} \label{eq:mse-32}
%MSE(\theta)&=\frac{1}{32}\sum_{t=1}^{32}\Big\{\big(R_{t}+\gamma\max_{a'}Q(S_{t+1},a';\theta')\big) \nonumber \\
%& \quad \quad\quad\quad\quad-Q(S_t,A_t;\theta)\Big\}^2
%\end{align}

After updating DNN features/parameters $\theta$, the agent also needs to update the target network parameters  $\theta'$ as follows (we call it soft update) \cite{Lillicrap_softupdate16}
\begin{align}
\theta'=\beta \theta + (1-\beta)\theta' \label{eq:soft-update}
\end{align}
where $\beta$ is update rate, $\beta \ll 1$.
%Note that iterative update $Q_{i+1}(S_t,A_t)=\mathbb{E} \big\{R_{t+1}\!+\!\gamma\max_{a'}Q_i(S_{t+1},a') | S_t, A_t \big\}$ converges to $Q^{\ast}(S_t,A_t)$ as $i \rightarrow \infty$. Instead, 

The explanation for why experience replay and target network mechanisms can improve algorithm stability has been given in \cite{Mnih_Nature15}.

\textbf{Optimal Action Policy:} Ideally, after the agent is trained, it will reach good estimate of the optimal Q-values and learn the optimal action policy accordingly. In reality, however, the agent may not learn good estimate of those optimal Q-values, because the agent has only experienced limited intersection states so far, not the overall state space, thus Q-values for states not experienced may not be well estimated. Moreover the state space itself may be changing continuously, making current estimated Q-values out of date. Therefore, the agent always faces a trade-off problem: whether to exploit already learned Q-values (which may not be accurate or out of date) and select the action with the greatest Q-value; or to explore other possible actions to improve Q-values estimate and finally improve action policy. We adopt a simple yet effective trade-off method, $\epsilon$-greedy method. Following this method, the agent selects the action with the current greatest estimated Q-value with probability $1-\epsilon$ (exploitation) and randomly selects one action with probability $\epsilon$ (exploration) at each time step.

%Instead of solving $(\ref{eq:recursive-q-value})$ directly, we resort to approximating those optimal Q-values $Q^{\ast}(s,a)$ iteratively as follows.
%\begin{align} \label{eq:q-learning}
%&\underbrace{Q(S_t,A_t)}_\text{new estimate} \leftarrow \underbrace{Q(S_t,A_t)}_\text{old estimate}\!+\alpha\big\{\underbrace{R_{t+1}\!+\!\gamma\max_{a}Q(S_{t+1},a)}_\text{target}\!-\!\underbrace{Q(S_t,A_t)}_\text{old estimate} \big\} \nonumber \\
%& \quad \quad \quad\quad\quad \text{for all} \, S_t \in \mathcal{S}, A_t \in \mathcal{A}
%\end{align}
%where $\alpha$ is a step-size parameter. 

%%%%%%%%%%%%%%%%%%%%%%%%%%%%%%%%%%%%%%%%%%%%%%%%%%%%%%%%%%%%%%%%%%%%%%%%%%%%%%%%%%%%%%%%%%
%%%%%%%%%%%%%%%%%%%%%%%%%%%%%%%%%%%%%%%%%%%%%%%%%%%%%%%%%%%%%%%%%%%%%%%%%%%%%%%%%%%%%%%%%%
\section{Simulation Evaluation} \label{section:ev}
In this section we first verify our deep reinforcement learning algorithm by simulations in terms of vehicle staying time, vehicle delay and algorithm stability, we then compare the vehicle delay of our algorithm to another two popular traffic signal control algorithms.

\subsection{Simulation Settings} \label{subsection:ss}
To simulate intersection traffic and traffic signal control, we use one popular open source simulator: Simulation of Urban MObility (SUMO) \cite{Krajzewicz_SUMO12}. Detailed simulation settings are as follows.

\textbf{Intersection:} Consider an intersection of four ways, each road with four lanes as shown in Fig. \ref{fig:intersection}. Set road length to be $500$ meters, road segment $l$ to be $160$ meters, cell length $c$ to be $8$ meters, road speed limit to be $19.444$ m/s (i.e., $70$ km/h), vehicle length to be $5$ meters, minimum gap between vehicles to be $2.5$ meters.

\textbf{Traffic Route:} All possible traffic routes at the intersection are summarized in Table \ref{table:routes}.
%Routes include vehicles going straight from west to east (WE) and turning left from west to north (WN), vehicles going straight from east to west (EW) and turning left from east to south (ES); similarly, vehicles going straight from north to south (NS) and turning left from north to east (NE), vehicles going straight from south to north (SN) and turning left from south to west (SW).

\begin{table}[t]
\caption{Traffic Routes}
\label{table:routes}
\centering
\begin{tabular}{ |l| l| }
\hline
\textbf{Route} & \textbf{Description} \\ \hline
$06$ &  going straight from road $0$ to road $6$ \\ \hline
$07$ &  turning left from road $0$ to road $7$ \\ \hline
$24$ &  going straight from road $2$ to road $4$ \\ \hline
$25$ &  turning left from road $2$ to road $5$ \\ \hline
$35$ &  going straight from road $3$ to road $5$ \\ \hline
$36$ &  turning left from road $3$ to road $6$ \\ \hline
$17$ &   going straight from road $1$ to road $7$ \\ \hline
$14$ &  turning left from road $1$ to road $4$ \\ \hline
\end{tabular}
\end{table}

\textbf{Traffic Arrival Process:} Vehicles arrive at road entrances randomly and select a route in advance. All arrivals follow the same Bernoulli process (an approximation to Poisson process) but with different rates $P_{ij}$, where $ij$ is route index, $i \in \{0, 1,2,3\}, j \in \{4,5,6,7\}$. For example, a vehicle following route $06$ will arrive at entrance of road $0$ with probability $P_{06}$ each second. To simulate heterogeneous traffic demands, we set roads $0,2$ to be busy roads and roads $1,3$ to be less busy roads. Specifically, $P_{06}=1/5, P_{07}=1/20, P_{24}=1/5, P_{25}=1/20, P_{35}=1/10, P_{36}=1/20, P_{17}=1/10, P_{14}=1/20$. All vehicles enter one road from random lanes.

\textbf{Traffic Signal Timing:} Rules for actuating traffic signals have been introduced in Agent Action of Section \ref{section:pf} and examples are given in Fig. \ref{fig:timeline} and Fig. \ref{fig:signaltiming}. Here, we set green light interval $\tau_g=10$ seconds and yellow light interval $\tau_y=6$ seconds.

\textbf{Agent Parameters:} The agent is trained for $N=2000$ episodes. Each episode corresponds to traffic of $1.5$ hours. For $\epsilon$-greedy method in Algorithm \ref{algorithm:drl}, parameter $\epsilon$ is set to be $0.1$ for all $2000$ episodes. Set discount factor $\gamma=0.95$, update rate $\beta=0.001$, learning rate of RMSProp algorithm to be $0.0002$ and capacity of replay memory to store data for $200$ episodes.

\textbf{Simulation data processing:} Define the delay of a vehicle at an intersection as the time interval (in seconds) from the time the vehicle enters one road of the intersection to the time it passes through/leaves the intersection. From the definition, we know that vehicle staying time is closely related to vehicle delay. During simulations, we record two types of data into separate files for all episodes: the sum of staying time of all vehicles at the intersection at every second and the delay of vehicles at each separate road $0, 1,2,3$. After collecting these data, we calculate their average values for each episode.

%\begin{table*}[!ht]
%\caption{Simulation settings}
%\label{table:settings}
%\centering
%\begin{tabular}{ |l| l| }
%\hline
%\textbf{Term} & \textbf{Setting details} \\ \hline
%intersection topology & four ways, each road with four lanes as shown in Fig. \ref{fig:intersection} \\ \hline
%traffic arrival &  arrival process for each road follows a Bernoulli process \\ \hline
%traffic route &   \\ \hline
%traffic signal timing & $\tau_g=3$ seconds, $\tau_y=6$ seconds \\ \hline
%\end{tabular}
%\end{table*}

\subsection{Simulation Results}
First, we examine simulation data to show that our algorithm indeed learns good action policy (i.e., traffic signal control policy) that effectively reduces vehicle staying time , thus reducing vehicle delay and traffic congestion, and that our algorithm is stable in making control decisions, i.e., not oscillating between good and bad action policies or even diverging to bad action policies. 

\begin{figure}[t]
\centering
\includegraphics[width=3.8in]{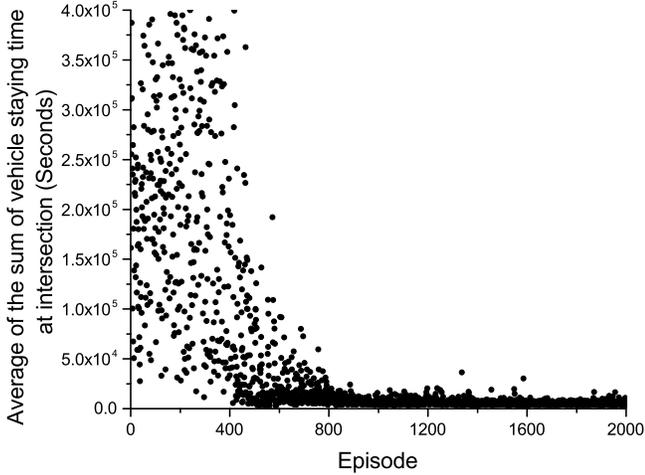}
\caption{Average of the sum of vehicle staying time at the intersection.}
\label{fig:softupdate}
\end{figure}

The average values for the sum of staying time of all vehicles at the intersection are presented in Fig. \ref{fig:softupdate}. From this figure, we can see that the average of the sum of vehicle staying time decreases rapidly as the agent is trained for more episodes and finally reduces to some small values, indicating that the agent does learn good action policy from training. We can also see that after $800$ episodes, average vehicle staying time keeps stable at small values, indicating that our algorithm converges to good action policy and algorithm stabilizing mechanisms, experience replay and target network, work effectively.

\begin{figure*}[!th]
    \centering
    {
    \subfloat[Road $0$]
    {\includegraphics[width=3.5in]{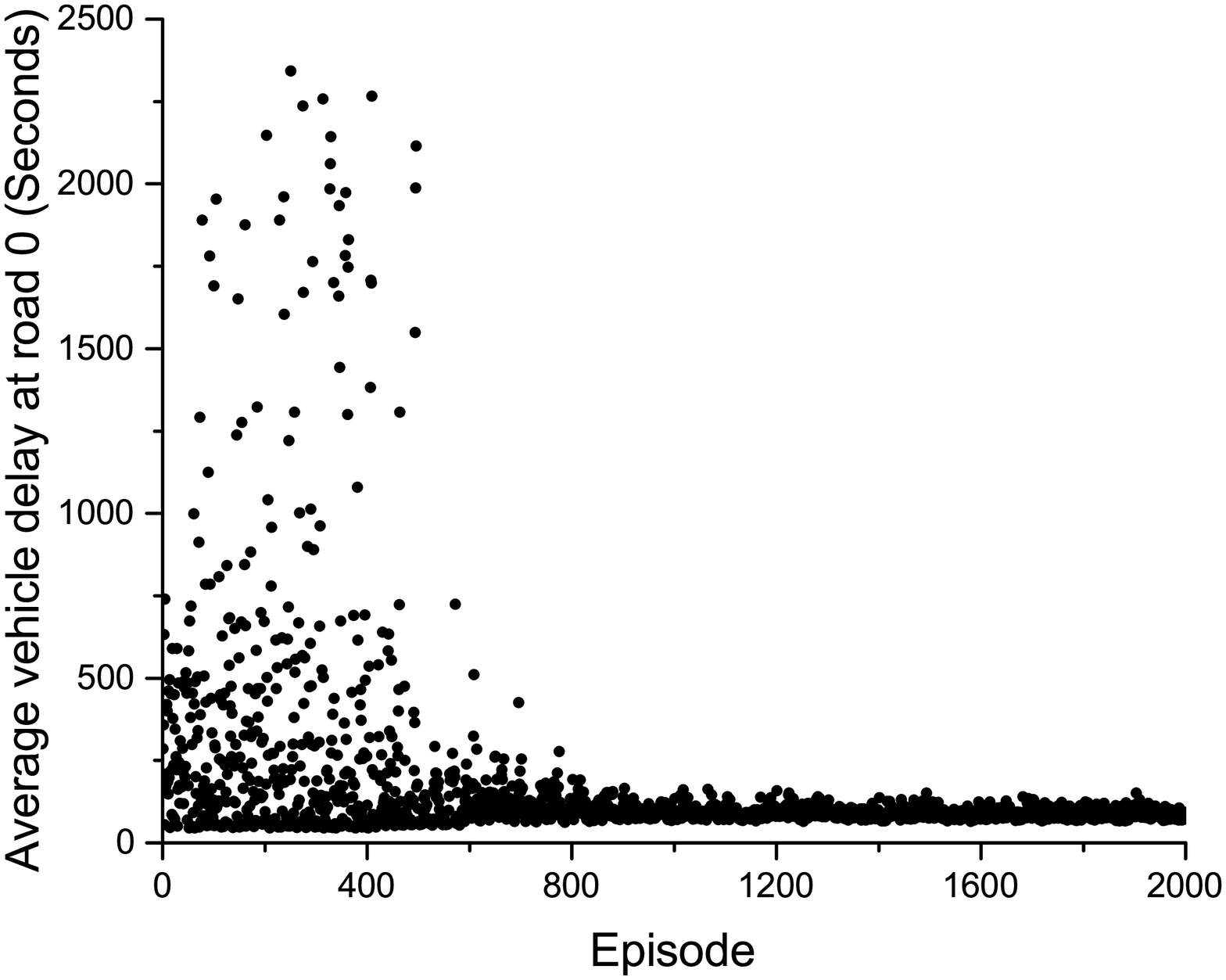} \label{fig:road0}}
    \subfloat[Road $1$]
    {\includegraphics[width=3.5in]{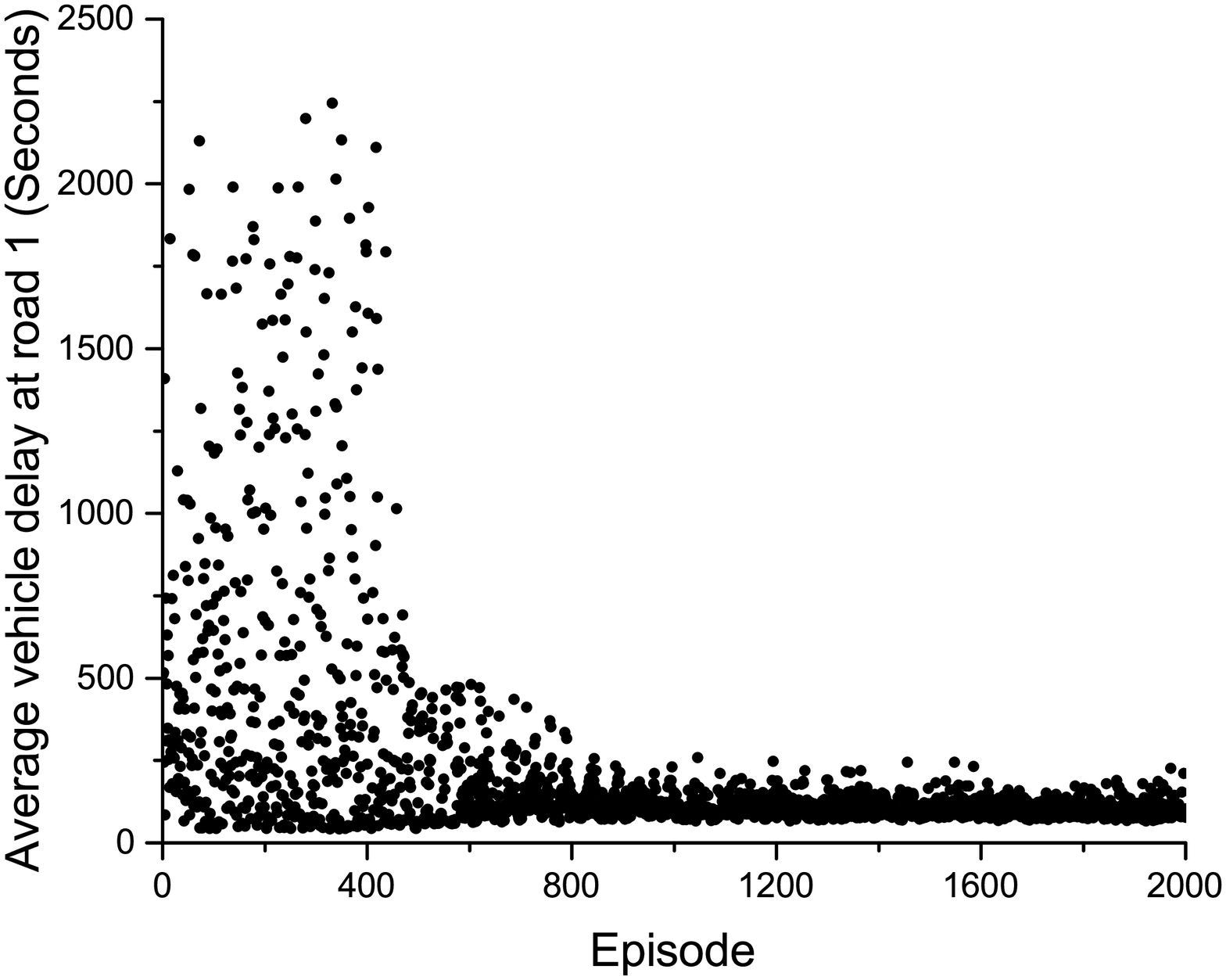} \label{fig:road1}}
    }
    {
    \subfloat[Road $2$]
    {\includegraphics[width=3.5in]{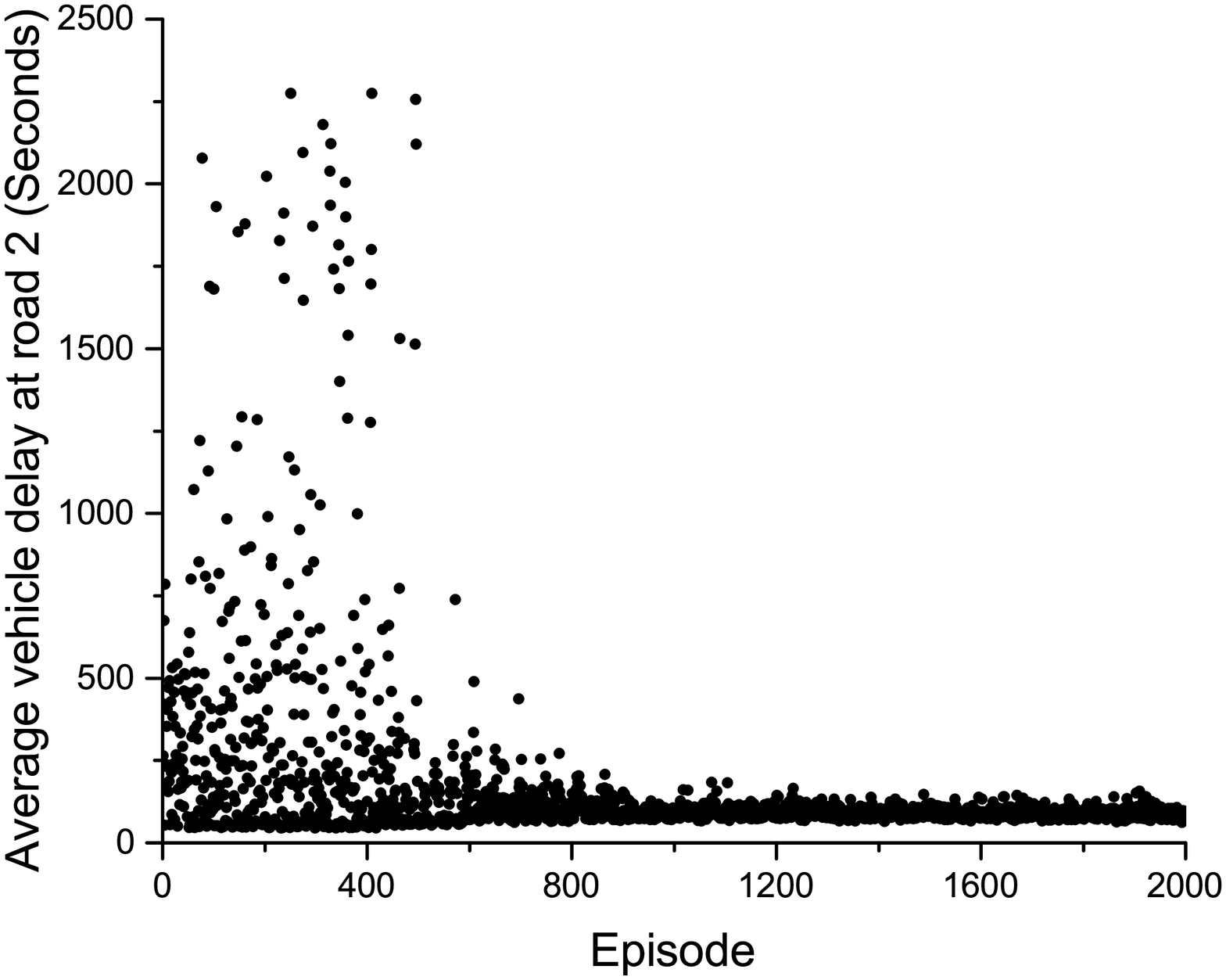} \label{fig:road2}}
    \subfloat[Road $3$]
    {\includegraphics[width=3.5in]{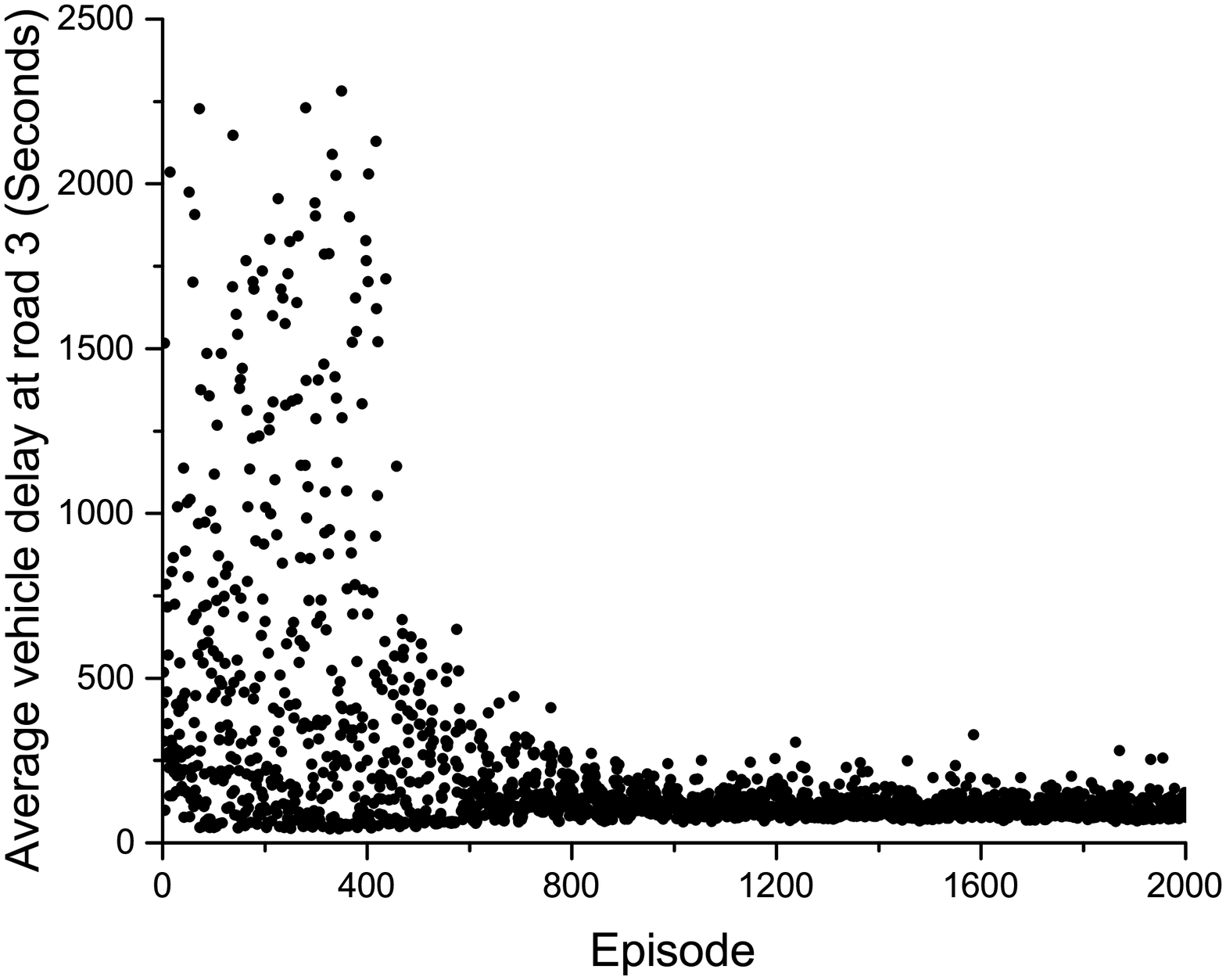} \label{fig:road3}}
    }
    \caption{Average vehicle delay for separate roads at the intersection.}
    \label{fig:delayroads}
\end{figure*}

The average values for delay of vehicles at each separate road are presented in Fig. \ref{fig:delayroads}. From this figure we see that average vehicle delay at each road is reduced greatly as the agent is trained for more episodes, indicating that our algorithm achieves adaptive and efficient traffic signal control. After the agent learns good action policy, average vehicle delay reduces to small values (around $90.5$ seconds for road $0$, $107.2$ seconds for road $1$, $91.5$ seconds for road $2$ and $109.4$ seconds for road $3$) and stays stable thereafter. From these stable values, we also know that our algorithm learns a fair policy: average vehicle delay for roads with different vehicle arrival rates does not differ too much. This is because long vehicle staying time, thus vehicle delay, at any road leads penalty to the agent (see ($\ref{eq:reward}$)), causing the agent to adjust its action policy accordingly. 

%Similar phenomenon can also be seen from the average values of the number of vehicles at each separate road presented in Fig. \ref{fig:vnumroads}.

\begin{figure*}[!th] 
    \centering
    {
    \subfloat[Road $0$]
    {\includegraphics[width=3.5in]{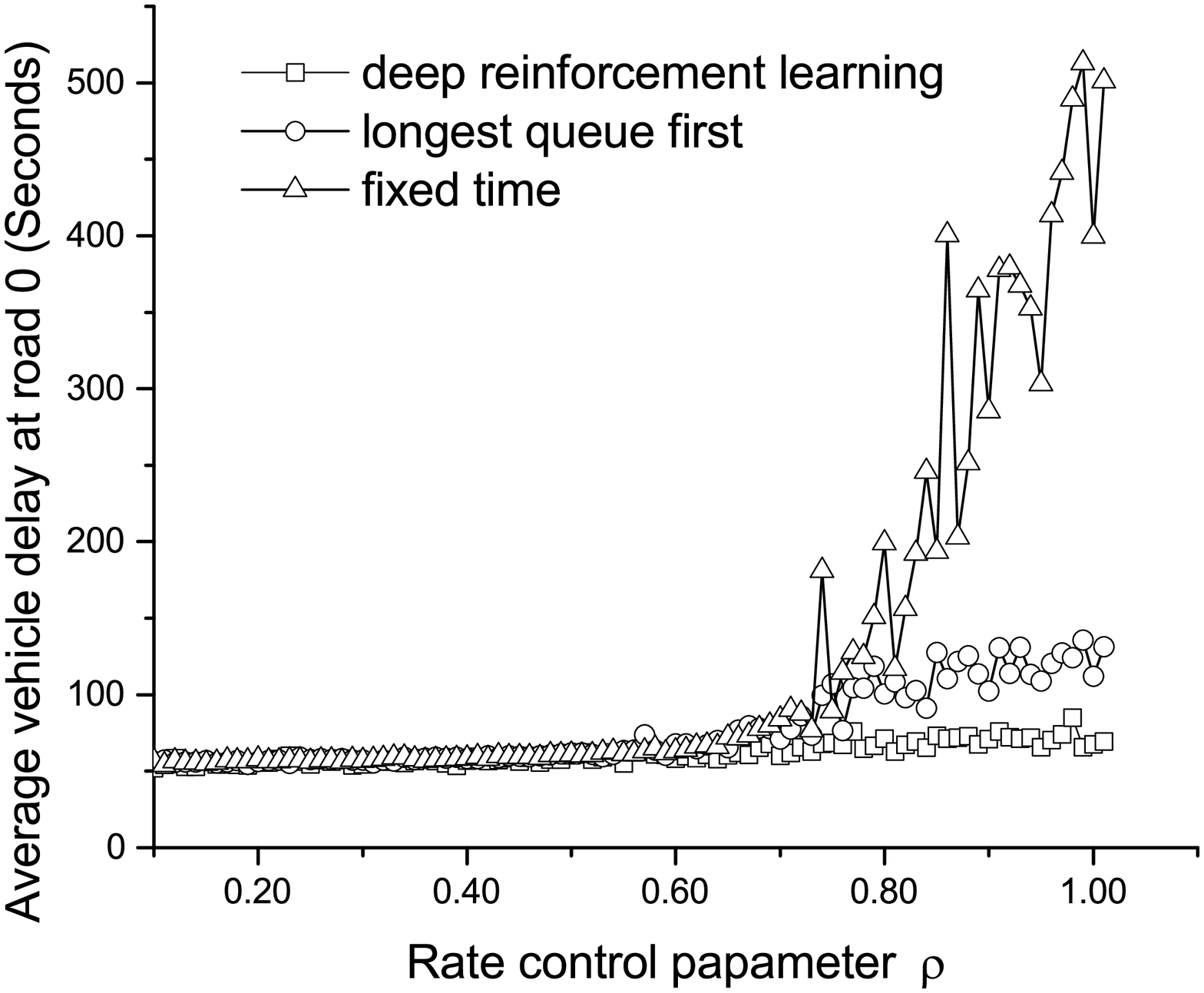} \label{fig:delayroad0com}}
    \subfloat[Road $1$]
    {\includegraphics[width=3.5in]{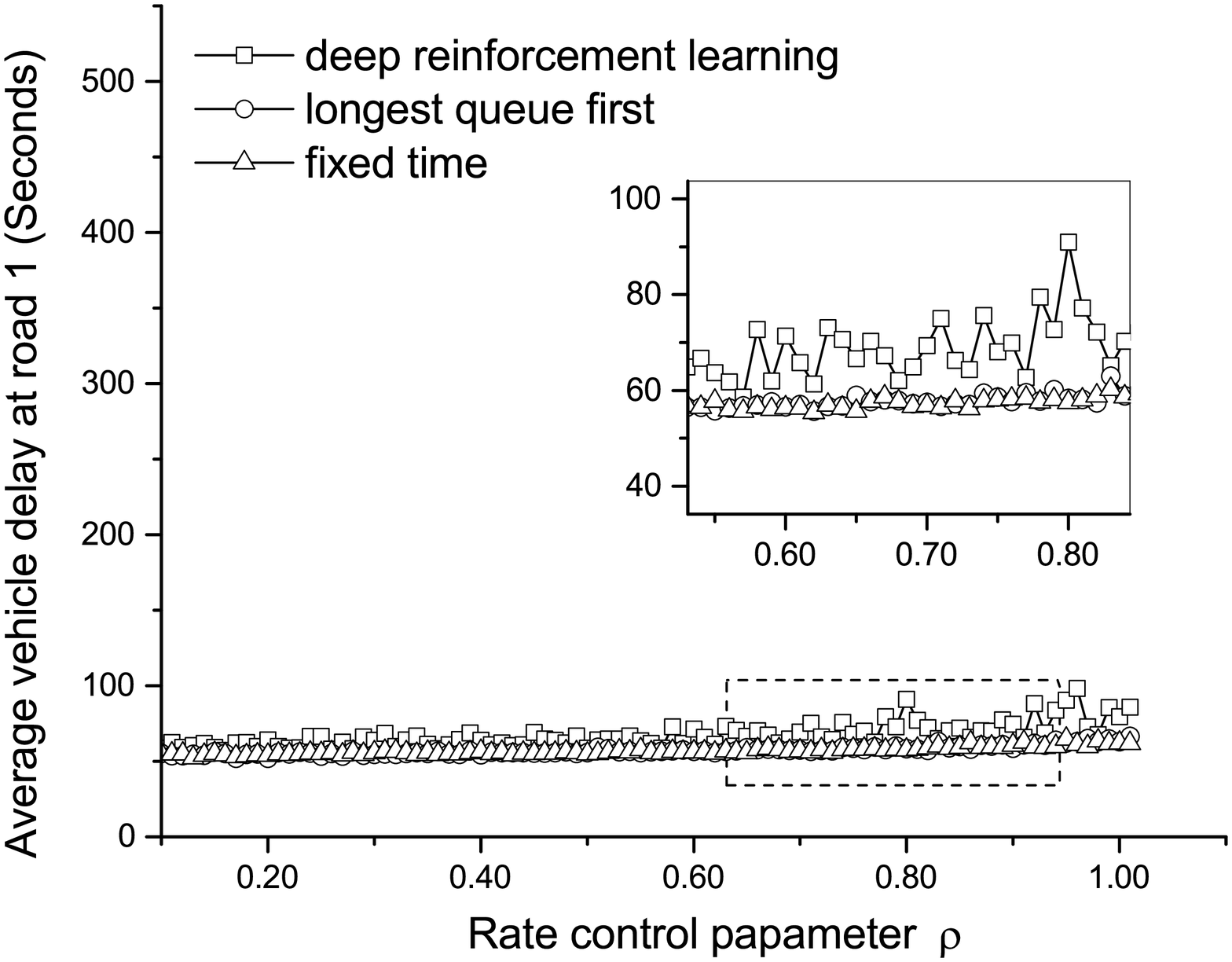} \label{fig:delayroad1com}}
    }
    {
    \subfloat[Road $2$]
    {\includegraphics[width=3.5in]{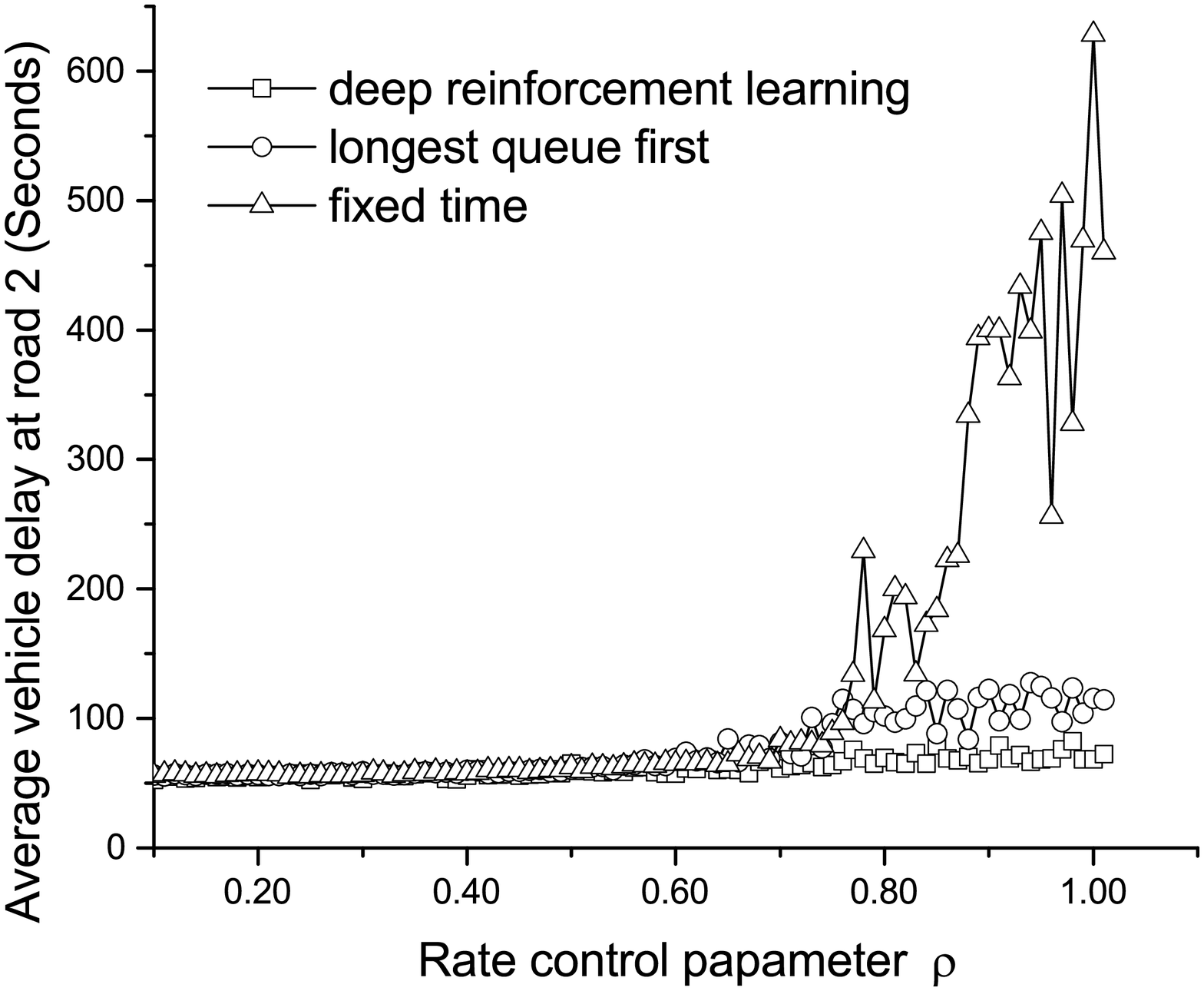} \label{fig:delayroad2com}}
    \subfloat[Road $3$]
    {\includegraphics[width=3.5in]{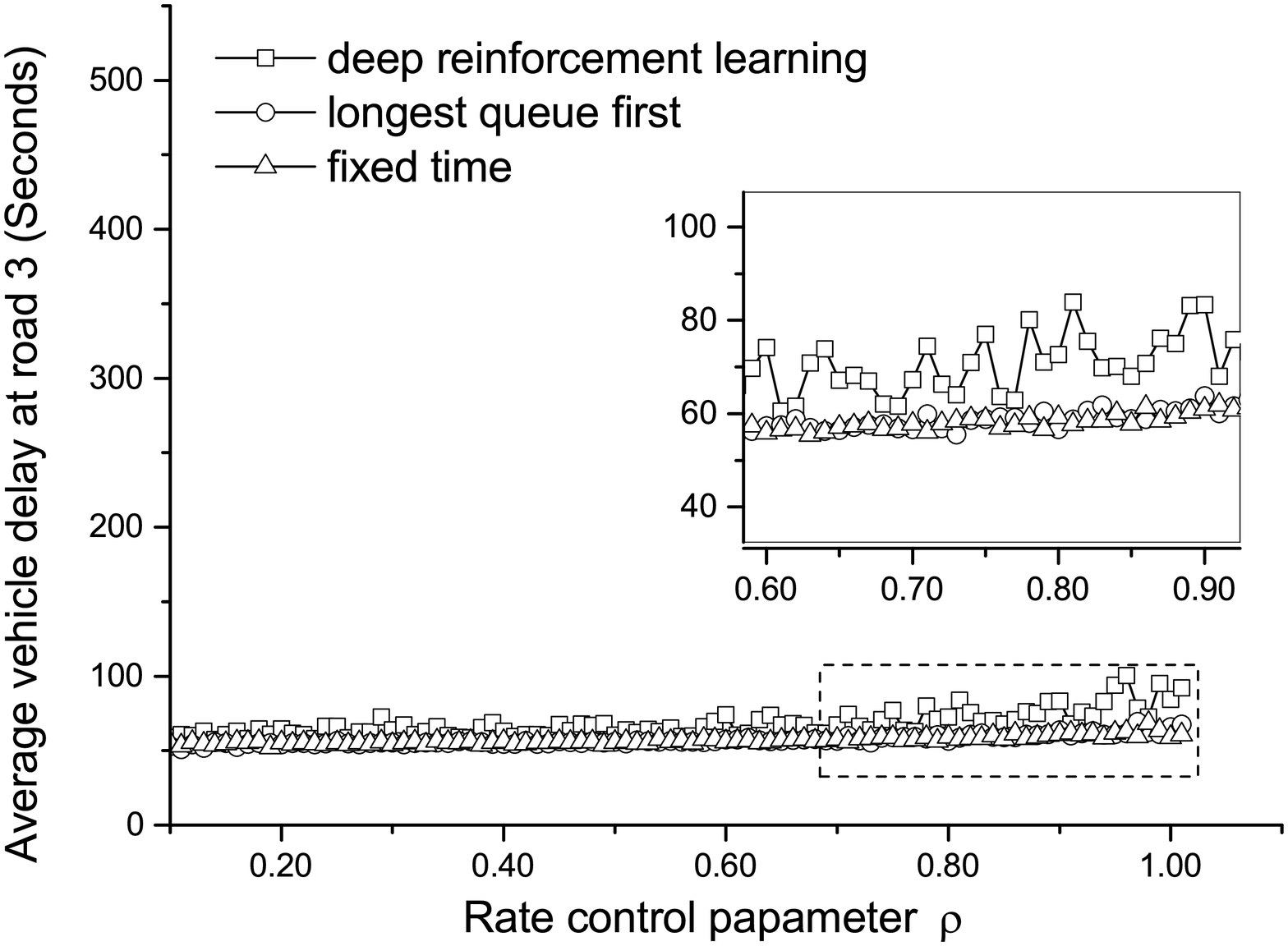} \label{fig:delayroad3com}}
    }
    \caption{Average vehicle delay for separate roads at the intersection under different traffic signal control algorithms.}
    \label{fig:delayroadsrho}
\end{figure*}

Next, we compare the vehicle delay performance of our algorithm to that of another two popular traffic signal control algorithms, longest queue first algorithm (turning on green lights for eligible traffic with most queued vehicles) \cite{Wunderlich_ITS08} and fixed time control algorithm (turning on green lights for eligible traffic using a predetermined cycle), under the same simulation settings in \ref{subsection:ss}. However, we change vehicle arrival rates by a parameter $\rho$ as $\rho P_{ij}$, $0.1 \leq \rho \leq 1$, during simulation, where values of $P_{ij}$, $i \in \{0, 1,2,3\}, j \in \{4,5,6,7\}$, are given in Section \ref{subsection:ss}. Simulation results are summarized in Fig.\ref{fig:delayroadsrho}. 

From Fig. \ref{fig:delayroadsrho}, we can see that for busy roads $0,2$, the average vehicle delay of our deep reinforcement learning algorithm is the lowest all the time: up to $86\%$ reduction when compared to fixed time control algorithm and up to $47\%$ reduction when compared to longest queue first algorithm. As traffic demand increases (i.e., as $\rho$  increases), average vehicle delay of fixed time control algorithm increases exponentially. This is because fixed time control algorithm is blind thus not adaptable to real-time traffic demands. Longest queue first algorithm can adapt to real-time traffic demand somewhat. However it only considers halting vehicles in queues, vehicles not in queues but to come soon are ignored, which is also useful information for traffic signal control. Our algorithm considers real time traffic information of all relevant vehicles, therefore outperforms the other two algorithms. Another observation from Fig. \ref{fig:delayroadsrho}$(a)$ and Fig. \ref{fig:delayroadsrho}$(c)$ is that as traffic demand increases, the average vehicle delay of our algorithm increases only slightly, indicating that our algorithm indeed adapts to dynamic traffic demand to reduce traffic congestion. However, this comes at the cost of slight increase in average vehicle delay at less busier roads $1,3$, as shown in the zoomed in portions of Fig. \ref{fig:delayroadsrho}$(b)$ and Fig. \ref{fig:delayroadsrho}$(d)$.

%%%%%%%%%%%%%%%%%%%%%%%%%%%%%%%%%%%%%%%%%%%%%%%%%%%%%%%%%%%%%%%%%%%%%%%%%%%%%%%%%%%%%%%%%%
%%%%%%%%%%%%%%%%%%%%%%%%%%%%%%%%%%%%%%%%%%%%%%%%%%%%%%%%%%%%%%%%%%%%%%%%%%%%%%%%%%%%%%%%%%
\section{Related Work} \label{section:rw}
In this section, we review related work on adopting deep reinforcement learning for traffic signal control.

After formulating traffic signal control problem as a reinforcement learning problem, Li \textit{et al.} proposed to use deep stacked autoencoders (SAE) neural network to estimate the optimal Q-values \cite{Li_JAS06}, where the algorithm takes the number of queued vehicles as input and queue difference between west-east traffic and north-south traffic as reward. By simulation, they compared the performance of their deep reinforcement learning algorithm to that of conventional reinforcement learning algorithm (i.e., without deep neural network) for traffic signal control, and concluded that deep reinforcement learning algorithm can reduce average traffic delay by $14\%$. However, they did not detail how target network is used for Q-value estimation nor how target network parameters are updated, which is important for stabilizing algorithm. Furthermore, they simulated an uncommon intersection scenario, where turning left, turning right are not allowed and there is no yellow clearance time. Whether their algorithm works for realistic intersection remains unknown. Different from this work, our algorithm does not use human-crafted feature, vehicle queue length, but automatically extracts all useful features from raw traffic data. Our algorithm works effectively for realistic intersections.

Aiming at realistic intersection, Genders \textit{et al.} \cite{Genders16} also proposed a deep reinforcement learning algorithm to adaptively control traffic signals, where convolutional neural networks are used to approximate optimal Q-values. Their algorithm takes vehicle position matrix, vehicle speed matrix and latest traffic signal as input, change in cumulative vehicle delay as reward, and uses a target network to estimate target Q-values. Through simulations, they showed that their algorithm could effectively reduce cumulative vehicle delay and vehicle travel time at an intersection. However, a well known problem with deep reinforcement learning is algorithm instability due to the moving target problem as explained in \cite{Pol16}. The authors did not mention how to solve this problem, a major drawback of their work. Furthermore, they did not consider fair traffic signal control issues as they mentioned and their intersection model does not have left-turn waiting areas, which is a commonly adopted and efficient mechanism for reducing vehicle delay at an intersection. In comparison, our algorithm not only improves algorithm stability but also finds fair traffic signal control policy for common intersections with left-turn waiting areas.
%Furthermore, their model of traffic signal transition is not realistic, where their algorithm controls traffic signals for vehicles turning left independently from vehicles going straight and turning right. For example, their algorithm can . However, in reality 

Pol addressed the moving target problem of deep reinforcement learning for traffic signal control in \cite{Pol16} and proposed to use a separate target network to approximate the target Q-values. Specifically, they fix target network parameters $\theta'$ for $M$ time steps during training, however update DNN network parameters $\theta$ every time step and copy DNN parameters $\theta$ into target network parameters $\theta'$ every $M$ time steps (referred to as hard update). By simulation, they showed that algorithm stability is improved if $M$ is set to be a proper value neither small nor large. However, this proper value of $M$ cannot be easily found in practice. Moreover, they used inefficient method to represent vehicle position information, which results in great computation cost during training. Specifically, the author used a binary position matrix: one indicating the presence of a vehicle at a position and zero indicating the absence of a vehicle at that position. Instead of covering only the roads area relevant to traffic signal control, they set the binary matrix to cover a whole rectangular area around the intersection. Since vehicles cannot run at areas except roads, most entries of the binary matrix are zero and redundant, making the binary matrix inefficient. Differently, our algorithm solves moving target problem by softly updating target network parameters $\theta'$, not needing to find proper value of $M$. Moreover, our algorithm represents vehicle position information efficiently (vehicle position matrix only covers intersection roads) thus reducing training computation cost.

%Different from all above related work, our algorithm works effectively for realistic intersection with left-turn waiting areas, solves moving target problem by a separate target network while not needing to find proper value of $M$, represents vehicle location information efficiently thus reducing training computation cost and finds fair yet efficient traffic signal control policy.

%our method solves moving target, not sparse computation reduction, soft copy no need to find proper $M$,
%
%
%definition of feature:
%Work \cite{Genders16} did not solve the instability problem of deep reinforcement learning algorithm, as discovered both in this paper and work \cite{Pol16}.
%
%Manual feature engineering
%
%Automatic feature engineering

%%%%%%%%%%%%%%%%%%%%%%%%%%%%%%%%%%%%%%%%%%%%%%%%%%%%%%%%%%%%%%%%%%%%%%%%%%%%%%%%%%%%%%%%%%
%%%%%%%%%%%%%%%%%%%%%%%%%%%%%%%%%%%%%%%%%%%%%%%%%%%%%%%%%%%%%%%%%%%%%%%%%%%%%%%%%%%%%%%%%%
\section{Conclusion} \label{section:conclusion}
We proposed a deep reinforcement learning algorithm for adaptive traffic signal control to reduce traffic congestion. Our algorithm can automatically extract useful features from raw real-time traffic data, which uses deep convolutional neural network, and learn the optimal traffic signal control policy. By adopting experience replay and target network mechanisms, we improved algorithm stability in the sense that our algorithm converges to good traffic signal control policy. Simulation results showed that our algorithm significantly reduces vehicle delay when compared to another two popular algorithms, longest queue first algorithm and fixed time control algorithm, and that our algorithm learns a fair traffic signal control policy such that no vehicles at any road wait too long for passing through the intersection.

\bibliographystyle{IEEEtran}
\bibliography{reference}

% Generated by IEEEtran.bst, version: 1.13 (2008/09/30)
\begin{thebibliography}{10}
\providecommand{\url}[1]{#1}
\csname url@samestyle\endcsname
\providecommand{\newblock}{\relax}
\providecommand{\bibinfo}[2]{#2}
\providecommand{\BIBentrySTDinterwordspacing}{\spaceskip=0pt\relax}
\providecommand{\BIBentryALTinterwordstretchfactor}{4}
\providecommand{\BIBentryALTinterwordspacing}{\spaceskip=\fontdimen2\font plus
\BIBentryALTinterwordstretchfactor\fontdimen3\font minus
  \fontdimen4\font\relax}
\providecommand{\BIBforeignlanguage}[2]{{%
\expandafter\ifx\csname l@#1\endcsname\relax
\typeout{** WARNING: IEEEtran.bst: No hyphenation pattern has been}%
\typeout{** loaded for the language `#1'. Using the pattern for}%
\typeout{** the default language instead.}%
\else
\language=\csname l@#1\endcsname
\fi
#2}}
\providecommand{\BIBdecl}{\relax}
\BIBdecl

\bibitem{Zhao_SMC12}
D.~Zhao, Y.~Dai, and Z.~Zhang, ``Computational intelligence in urban traffic
  signal control: A survey,'' \emph{IEEE Transactions on Systems, Man, and
  Cybernetics, Part C (Applications and Reviews)}, vol.~42, no.~4, pp.
  485--494, July 2012.

\bibitem{Alsabaan_CST13}
M.~Alsabaan, W.~Alasmary, A.~Albasir, and K.~Naik, ``Vehicular networks for a
  greener environment: A survey,'' \emph{IEEE Communications Surveys \&
  Tutorials}, vol.~15, no.~3, pp. 1372--1388, Third Quarter 2013.

\bibitem{Zaidi_ITS16}
A.~A. Zaidi, B.~Kulcsár, and H.~Wymeersch, ``Back-pressure traffic signal
  control with fixed and adaptive routing for urban vehicular networks,''
  \emph{IEEE Transactions on Intelligent Transportation Systems}, vol.~17,
  no.~8, pp. 2134--2143, August 2016.

\bibitem{Gregoire_CNS15}
J.~Gregoire, X.~Qian, E.~Frazzoli, A.~de~La~Fortelle, and T.~Wongpiromsarn,
  ``Capacity-aware backpressure traffic signal control,'' \emph{IEEE
  Transactions on Control of Network Systems}, vol.~2, no.~2, pp. 164--173,
  June 2015.

\bibitem{LA_ITS11}
P.~LA and S.~Bhatnagar, ``Reinforcement learning with function approximation
  for traffic signal control,'' \emph{IEEE Transactions on Intelligent
  Transportation Systems}, vol.~12, no.~2, pp. 412--421, June 2011.

\bibitem{Yin_CDC15}
B.~Yin, M.~Dridi, and A.~E. Moudni, ``Approximate dynamic programming with
  recursive least-squares temporal difference learning for adaptive traffic
  signal control,'' in \emph{IEEE 54th Annual Conference on Decision and
  Control (CDC)}, 2015.

\bibitem{Arel_IET10}
I.~Arel, C.~Liu, T.~Urbanik, and A.~G. Kohls, ``Reinforcement learning-based
  multi-agent system for network traffic signal control,'' \emph{IET
  Intelligent Transport Systems}, vol.~4, no.~2, pp. 128--135, June 2010.

\bibitem{Mannion_chapter16}
P.~Mannion, J.~Duggan, and E.~Howley, \emph{An Experimental Review of
  Reinforcement Learning Algorithms for Adaptive Traffic Signal Control}.\hskip
  1em plus 0.5em minus 0.4em\relax Springer International Publishing, May 2016,
  ch. Autonomic Road Transport Support Systems, pp. 47--66.

\bibitem{Neely_PhD03}
M.~J. Neely, ``Dynamic power allocation and routing for satellite and wireless
  networks with time varying channels,'' Ph.D. dissertation, LIDS,
  Massachusetts Institute of Technology, Cambridge, MA, USA, 2003.

\bibitem{Sutton_RL98}
R.~S. Sutton and A.~G. Barto, \emph{Reinforcement Learning: An
  Introduction}.\hskip 1em plus 0.5em minus 0.4em\relax MIT Press, 1998.

\bibitem{Mnih_Nature15}
V.~Mnih, K.~Kavukcuoglu, D.~Silver, A.~A. Rusu, J.~Veness, M.~G. Bellemare,
  A.~Graves, M.~Riedmiller, A.~K. Fidjeland, G.~Ostrovski, S.~Petersen,
  C.~Beattie, A.~Sadik, I.~Antonoglou, H.~King, D.~Kumaran, D.~Wierstra,
  S.~Legg, and D.~Hassabis, ``Human-level control through deep reinforcement
  learning,'' \emph{Nature}, vol. 518, no. 7540, pp. 529--533, 2015.

\bibitem{Genders16}
W.~Genders and S.~Razavi, ``Using a deep reinforcement learning agent for
  traffic signal control,'' November 2016, [Online]. Available:
  https://arxiv.org/abs/1611.01142.

\bibitem{Hinton_RMSProp12}
T.~Tieleman and G.~Hinton, ``Lecture 6.5-rmsprop: Divide the gradient by a
  running average of its recent magnitude,'' \emph{COURSERA: Neural networks
  for machine learning}, vol.~4, no.~2, 2012.

\bibitem{Lillicrap_softupdate16}
T.~P. Lillicrap, J.~J. Hunt, A.~Pritzel, N.~Heess, T.~Erez, Y.~Tassa,
  D.~Silver, and D.~Wierstra, ``Continuous control with deep reinforcement
  learning,'' February 2016, [Online]. Available:
  https://arxiv.org/abs/1509.02971.

\bibitem{Krajzewicz_SUMO12}
D.~Krajzewicz, J.~Erdmann, M.~Behrisch, and L.~Bieker, ``Recent development and
  applications of sumo – simulation of urban mobility,'' \emph{International
  Journal On Advances in Systems and Measurements}, vol.~5, no. 3 \& 4, pp.
  128--138, December 2012.

\bibitem{Wunderlich_ITS08}
R.~Wunderlich, C.~Liu, and I.~Elhanany, ``A novel signal-scheduling algorithm
  with quality-of-service provisioning for an isolated intersection,''
  \emph{IEEE Transactions on Intelligent Transportation Systems}, vol.~9,
  no.~3, pp. 536--547, September 2008.

\bibitem{Li_JAS06}
L.~Li, Y.~Lv, and F.-Y. Wang, ``Traffic signal timing via deep reinforcement
  learning,'' \emph{IEEE/CAA Journal of Automatica Sinica}, vol.~3, no.~3, pp.
  247-- 254, July 2016.

\bibitem{Pol16}
E.~van~der Pol, ``Deep reinforcement learning for coordination in traffic light
  control,'' Master's thesis, University of Amsterdam, August 2016.

\end{thebibliography}

\end{document}